\documentclass[aps,twocolumn]{revtex4}%
\usepackage{amssymb}
\usepackage{amsfonts}
\usepackage{amsmath}
\usepackage{graphicx}
\usepackage[usenames]{color}%
\providecommand{\U}[1]{\protect\rule{.1in}{.1in}}
\begin{document}
\preprint{ }
\title[Leggett collective excitations]{Leggett collective excitations in a two-band Fermi superfluid at finite temperatures}
\author{S. N. Klimin}
\affiliation{{TQC, Universiteit Antwerpen, Universiteitsplein 1, B-2610 Antwerpen, Belgium}}
\author{H. Kurkjian}
\affiliation{{TQC, Universiteit Antwerpen, Universiteitsplein 1, B-2610 Antwerpen, Belgium}}
\author{J. Tempere}
\affiliation{{TQC, Universiteit Antwerpen, Universiteitsplein 1, B-2610 Antwerpen, Belgium}}
\affiliation{{Lyman Laboratory of Physics, Harvard University, USA}}

\begin{abstract}
The Leggett collective excitations for a two-band Fermi gas with $s$-wave
pairing and Josephson interband coupling in the BCS-BEC crossover at finite
temperatures are investigated within the Gaussian pair fluctuation approach.
Eigenfrequencies and damping factors for Leggett modes are determined in a
nonperturbative way, using the analytic continuation of the fluctuation
propagator through a branch cut in the complex frequency plane, as in
\emph{Phys. Rev. Lett. \textbf{122}, 093403 (2019)}. The treatment is
performed beyond the low-energy expansion, which is necessary when the
collective excitation energy reaches the pair-breaking continuum edge. The
results are applied in particular to cold atomic gases at the orbital Feshbach
resonance and in a regime far from BEC, which can be relevant for future experiments.

\end{abstract}
\date{\today}
\maketitle

\section{Introduction}

The collective excitations predicted by Leggett \cite{Leggett} are specific
for multiband neutral or charged superfluids. Physically, they describe
oscillations of the relative phase of different superfluid band-components. An
existence of this branch of collective excitations necessarily follows from
the simple reasoning. In a one-band BCS superfluid or superconductor, there
can exist two branches of collective excitations: the phononic (Goldstone)
branch of phase oscillations and the pair-breaking (Higgs) branch of amplitude
oscillations. In a two-band superfluid with a Josephson interband coupling,
there still exists a single gapless Goldstone branch, which corresponds to
oscillations of the common phase of the two superfluid band components. For
completeness, a branch of phase collective excitations must exist which
correspond to the motion of the \emph{relative} phase of the two components
and has a finite gap due to the Josephson coupling. These oscillations, by
definition, are absent in a one-band-component superfluid/superconducting
system. For cold Fermi gases which are not in the BCS regime, the picture is
more complicated, because eigenmodes of collective excitations are not purely
phase or amplitude, as obtained in the present work.

Leggett modes were experimentally detected in multiband superconductors
\cite{Blumberg,Ponomarev,Giorgianni}. The two-band superfluidity and Leggett
collective modes have been considered theoretically for superconducting
MgB$_{2}$ \cite{Maksimov} and for condensed Fermi gases in the BCS-BEC
crossover \cite{Iskin1,Iskin2}. It was not clear for a long time whether the
two-band superfluidity and Leggett collective excitations can be
experimentally achieved in cold atomic gases. The idea of the two-band
superfluid scenario based on the so-called orbital Feshbach resonance (OFR)
was theoretically proposed by R. Zhang \emph{et al}. \cite{Zhang2015}. OFR has
been later successfully realized in a condensed $^{173}$Yb gas
\cite{Pagano,Hofer}. This stimulated theoretical efforts to describe the
two-band superfluid states \cite{Iskin3,He2015} and Leggett modes
\cite{He2016,Zhang}. BCS-BEC crossover in a two-band atomic superfluid has
been studied at the mean-field level and with fluctuations pointing to unique
features of the two-component condensate in recent works
\cite{Tajima2019,Yerin2019}. The Leggett mode frequencies in cold Fermi gases
with OFR were first calculated in Ref. \cite{He2016} at zero temperature,
solving the Gaussian pair fluctuation propagator for undamped modes. The work
\cite{Zhang} reprersents an alternative method exploiting the density-density
response function. In the recent work \cite{Zou2018}, the massive Leggett mode
and gapless phonon mode of the two-band Fermi superfluid are studied for the
Sarma phase.

For nonzero temperatures, two-band superfluidity and Leggett modes in cold
Fermi gases were investigated earlier in Ref. \cite{Klimin2015} using
effective field theory (EFT), which assumes pair fields slowly varying in time
and space. The EFT is applicable only for low-energy excitations, which is not
always the case for Leggett modes. Moreover, it is inapplicable for energies
in the vicinity of the pair-breaking continuum edge \cite{Kurkjian2019}.
Therefore the study of Leggett collective excitations at nonzero temperature
beyond the low-energy approximation is timely and relevant.

In the present work, we consider Leggett collective modes in a two-band Fermi
superfluid using the Gaussian pair fluctuation (GPF) effective action for a
two-band system derived in Ref. \cite{Klimin2015} within the path integral
formalism. The frequencies and the damping factors for Leggett modes are
determined through complex poles of the fluctuation propagator, similarly to
Refs. \cite{Kurkjian2019,Arxiv}, where this method has been applied to
pair-breaking and phononic collective excitations in ultracold Fermi gases. In
the literature, the damping factor of collective excitations is frequently
determined through a second-order perturbation expression with a real
eigenfrequency. This is well substantiated only when damping is relatively
small with respect to the frequency. Here, the complex poles are found as
proposed by P. Nozi\`{e}res \cite{Nozieres}, so that the real and imaginary
part of the complex excitation frequency are determined mutually consistently
and beyond the aforesaid perturbation approach. In this context, the present
calculation can be classified as nonperturbative. The GPF approach is also not
restricted to the weak interband coupling regime and is valid in the whole
temperature range below the transition temperature $T_{c}$.

The spectra of Leggett modes are analyzed as a function of the coupling
parameters, temperature, and the detuning factor $\delta$, which characterizes
the band offset between the two bands. Here, numeric results are presented for
the long-wavelength limit, focusing first on a superfluid Fermi gas in the BEC
regime, which is relevant for the recent experimental achievements
\cite{Pagano,Hofer}. It was shown that the whole BCS-BEC crossover can be
realized for superfluid Fermi gases with OFR \cite{He2016}. Consequently, we
also consider Leggett collective modes in the BCS-BEC crossover regime.

\section{GPF method for two bands}

We use the model fermionic action with $s$-wave intraband and interband
pairing channels proposed in Ref. \cite{Klimin2015}, expressed in Grassmann
variables $\left\{  \bar{\psi}_{\sigma,j},\psi_{\sigma,j}\right\}  $, where
$\sigma$ and $j$ are, respectively, the spin-component and band-component
indices. Assuming $\hbar=1$, this action, with the inverse temperature $\beta
$, reads:%
\begin{align}
S  &  =\int_{0}^{\beta}d\tau\int d\mathbf{r}\left[  \sum_{j=1,2}\sum
_{\sigma=\uparrow,\downarrow}\bar{\psi}_{\sigma,j}\left(  \frac{\partial
}{\partial\tau}\right.  \right. \nonumber\\
&  \left.  \left.  -\frac{\nabla_{\mathbf{r}}^{2}}{2m}-\mu_{\sigma,j}\right)
\psi_{\sigma,j}+U\right]  . \label{SFerm}%
\end{align}
The chemical potentials $\mu_{\sigma,j}$ can be, in general, different for the
two band and two spin components. The interaction term $U\left(
\mathbf{r},\tau\right)  $ describes the contact interactions between fermions
for both intraband and interband contact interactions:%
\begin{align}
U  &  =%
%TCIMACRO{\tsum \nolimits_{j=1,2}}%
%BeginExpansion
{\textstyle\sum\nolimits_{j=1,2}}
%EndExpansion
g_{j}\bar{\psi}_{\uparrow,j}\bar{\psi}_{\downarrow,j}\psi_{\downarrow,j}%
\psi_{\uparrow,j}\nonumber\\
&  +g_{3}\left(  \bar{\psi}_{\uparrow,1}\psi_{\uparrow,1}\bar{\psi
}_{\downarrow,2}\psi_{\downarrow,2}+\bar{\psi}_{\downarrow,1}\psi
_{\downarrow,1}\bar{\psi}_{\uparrow,2}\psi_{\uparrow,2}\right) \nonumber\\
&  +g_{4}\left(  \bar{\psi}_{\uparrow,1}\psi_{\uparrow,1}\bar{\psi}%
_{\uparrow,2}\psi_{\uparrow,2}+\bar{\psi}_{\downarrow,1}\psi_{\downarrow
,1}\bar{\psi}_{\downarrow,2}\psi_{\downarrow,2}\right)  . \label{U}%
\end{align}
The contributions for the two intraband Cooper pairing channels are with
$g_{1},g_{2}$, while those with $g_{3},g_{4}$ are related to the interband
fermion-fermion scattering. After the Hubbard-Stratonovich transformation with
auxiliary bosonic fields as described in \cite{Klimin2015} we arrive at the
effective bosonic action,%
\begin{equation}
S_{eff}^{\left(  2b\right)  }=\sum_{j=1,2}S_{eff}^{\left(  j\right)  }%
-\int_{0}^{\beta}d\tau\int d\mathbf{r~}\frac{m\gamma}{4\pi}\left(  \bar{\Psi
}_{1}\Psi_{2}+\bar{\Psi}_{2}\Psi_{1}\right)  , \label{S2B}%
\end{equation}
with the Josephson interband coupling strength $\gamma=\left(  8\pi/m\right)
\left(  1/g_{3}-1/g_{4}\right)  $. In should be noted that the Josephson
coupling refers to the exchange of Cooper pairs between the two bands. Hence
it is a coupling term between two partial condensates of the system, as
distinct from the cross-band pairing, i. e., formation of interband Cooper
pairs, which is not considered here.

The effective action $S_{eff}^{\left(  j\right)  }$ for each band depends on
the bosonic pair fields $\left\{  \bar{\Psi}_{j},\Psi_{j}\right\}  $ in the
same way as in the one-band theory, e. g., \cite{SDM1993}:%
\begin{equation}
S_{eff}^{\left(  j\right)  }=-\operatorname{Tr}\ln\left[  -\mathbb{G}_{j}%
^{-1}\right]  -\int_{0}^{\beta}d\tau\int d\mathbf{r}\frac{1}{g_{j}}\bar{\Psi
}_{j}\Psi_{j}, \label{Seff}%
\end{equation}
with the inverse Nambu tensor:%
\begin{equation}
-\mathbb{G}_{j}^{-1}=\left(
\begin{array}
[c]{cc}%
\frac{\partial}{\partial\tau}-\frac{\nabla_{\mathbf{r}}^{2}}{2m}-\mu
_{\uparrow,j} & -\Psi\\
-\bar{\Psi} & \frac{\partial}{\partial\tau}+\frac{\nabla_{\mathbf{r}}^{2}}%
{2m}+\mu_{\downarrow,j}%
\end{array}
\right)  . \label{G}%
\end{equation}

Collective excitations are small Gaussian pair fluctuations $\left\{
\bar{\varphi}_{j},\varphi_{j}\right\}  $ about the saddle-point gap solution
$\left\{  \bar{\Delta}_{j},\Delta_{j}\right\}  $,
\begin{equation}
\Psi_{j}=\Delta_{j}+\varphi_{j},\quad\bar{\Psi}_{j}=\bar{\Delta}_{j}%
+\bar{\varphi}_{j}.
\end{equation}
The saddle-point gap is determined for a two-band system by the coupled set of
gap equations. The gap equations are determined using the standard
renormalization of the coupling constants for the contact interaction
\cite{SDM1993} which removes ultraviolet divergences and results in $s$-wave
scattering lengths $a_{j}$:%
\begin{equation}
\frac{1}{g_{j}}=\frac{m}{4\pi a_{j}}-\int\frac{d^{3}k}{\left(  2\pi\right)
^{3}}\frac{m}{k^{2}}. \label{renorm}%
\end{equation}
The coupled gap equations are then given by:%
\begin{align}
&  \left[  \int\frac{d\mathbf{k}}{\left(  2\pi\right)  ^{3}}\left(
\frac{X\left(  E_{\mathbf{k}}^{\left(  j\right)  }\right)  }{2E_{\mathbf{k}%
}^{\left(  j\right)  }}-\frac{m}{k^{2}}\right)  +\frac{m}{4\pi a_{j}}\right]
\Delta_{j}\nonumber\\
&  +\frac{m\gamma}{4\pi}\Delta_{3-j}=0, \label{gapeqs}%
\end{align}
with the function
\begin{equation}
X\left(  E_{\mathbf{k}}^{\left(  j\right)  }\right)  =\frac{\sinh\left(  \beta
E_{\mathbf{k}}^{\left(  j\right)  }\right)  }{\cosh\left(  \beta
E_{\mathbf{k}}^{\left(  j\right)  }\right)  +\cosh\left(  \beta\zeta
_{j}\right)  } \label{X}%
\end{equation}
depending on the Bogoliubov excitation energy $E_{\mathbf{k}}^{\left(
j\right)  }=\sqrt{\left(  \xi_{\mathbf{k}}^{\left(  j\right)  }\right)
+\Delta_{j}^{2}}$, where $\xi_{\mathbf{k}}^{\left(  j\right)  }=k^{2}%
/2m-\mu_{j}$ is the free-fermion energy. We have denoted here $\mu_{j}=\left(
\mu_{\uparrow,j}+\mu_{\downarrow,j}\right)  /2$ and $\zeta_{j}=\left(
\mu_{\uparrow,j}-\mu_{\downarrow,j}\right)  /2$. When setting equal scattering
lengths for the two band components, the effective action derived in
\cite{Klimin2015} is reduced to the model Hamiltonian for OFR of Ref.
\cite{Zhang}.

The GPF action is a quadratic form of fluctuation coordinates in the two-band
superfluid,%
\begin{equation}
S_{GPF}^{\left(  2b\right)  }=\frac{1}{2}\sum_{\mathbf{q},n}\bar{\varphi
}_{\mathbf{q},n}\mathbb{M}_{2b}\left(  \mathbf{q},i\Omega_{n}\right)
\varphi_{\mathbf{q},n}. \label{SGPF}%
\end{equation}
Here, $\varphi_{\mathbf{q},n}$ are 4-dimensional Nambu vectors:%
\begin{equation}
\varphi_{\mathbf{q},n}=\left(
\begin{array}
[c]{c}%
\varphi_{\mathbf{q},n}^{\left(  1\right)  }\\
\bar{\varphi}_{-\mathbf{q},-n}^{\left(  1\right)  }\\
\varphi_{\mathbf{q},n}^{\left(  2\right)  }\\
\bar{\varphi}_{-\mathbf{q},-n}^{\left(  2\right)  }%
\end{array}
\right)  , \label{Nambu}%
\end{equation}
where the superscripts indicate the band components. The inverse GPF
propagator for the two-band system $\mathbb{M}_{2b}\left(  \mathbf{q}%
,i\Omega_{n}\right)  $ depending on the pair momentum $\mathbf{q}$ and the
Matsubara frequency $\Omega_{n}=2\pi n/\beta$ can be written in the form%
\begin{equation}
\mathbb{M}_{2b}=\left(
\begin{array}
[c]{cc}%
\mathbb{M}^{\left(  1\right)  }+\varkappa\frac{\Delta_{2}}{\Delta_{1}%
}\mathbb{I} & -\varkappa\mathbb{I}\\
-\varkappa\mathbb{I} & \mathbb{M}^{\left(  2\right)  }+\varkappa\frac
{\Delta_{1}}{\Delta_{2}}\cdot\mathbb{I}%
\end{array}
\right)  , \label{M2b}%
\end{equation}
where $\varkappa\equiv m\gamma/\left(  4\pi\right)  $, $\mathbb{I}$ is the
unit $2\times2$ matrix, and $\mathbb{M}^{\left(  j\right)  }\left(
q,i\Omega_{n}\right)  $ is the inverse GPF propagator for the $j$-th band
component, with the matrix elements \cite{Engelbrecht,Diener2008,PRA2008} (for
simplicity, we drop here the band-component index $j$):%
\begin{align}
&  \left.  M_{1,1}\left(  \mathbf{q},i\Omega_{n}\right)  =M_{2,2}\left(
-\mathbf{q},-i\Omega_{n}\right)  \right. \nonumber\\
&  {=\int\frac{d^{3}k}{\left(  2\pi\right)  ^{3}}}\frac{X\left(
E_{\mathbf{k}}\right)  }{2E_{\mathbf{k}}}{\left[  1+\frac{1}{2E_{\mathbf{k}%
+\mathbf{q}}}\right.  }\nonumber\\
&  \times\left(  \frac{\left(  \xi_{\mathbf{k}}+E_{\mathbf{k}}\right)  \left(
E_{\mathbf{k}+\mathbf{q}}+\xi_{\mathbf{k}+\mathbf{q}}\right)  }{i\Omega
_{n}-E_{\mathbf{k}}-E_{\mathbf{k}+\mathbf{q}}}-\frac{\left(  \xi_{\mathbf{k}%
}+E_{\mathbf{k}}\right)  \left(  \xi_{\mathbf{k}+\mathbf{q}}-E_{\mathbf{k}%
+\mathbf{q}}\right)  }{i\Omega_{n}-E_{\mathbf{k}}+E_{\mathbf{k}+\mathbf{q}}%
}\right. \nonumber\\
&  \left.  \left.  +\frac{\left(  \xi_{\mathbf{k}}-E_{\mathbf{k}}\right)
\left(  \xi_{\mathbf{k}+\mathbf{q}}+E_{\mathbf{k}+\mathbf{q}}\right)
}{i\Omega_{n}+E_{\mathbf{k}}-E_{\mathbf{k}+\mathbf{q}}}-\frac{\left(
\xi_{\mathbf{k}}-E_{\mathbf{k}}\right)  \left(  \xi_{\mathbf{k}+\mathbf{q}%
}-E_{\mathbf{k}+\mathbf{q}}\right)  }{i\Omega_{n}+E_{\mathbf{k}}%
+E_{\mathbf{k}+\mathbf{q}}}\right)  \right]  , \label{M11}%
\end{align}
and%
\begin{align}
&  \left.  M_{1,2}\left(  \mathbf{q},i\Omega_{n}\right)  =M_{2,1}\left(
-\mathbf{q},-i\Omega_{n}\right)  \right. \nonumber\\
&  =-\Delta^{2}\int\frac{d^{3}k}{\left(  2\pi\right)  ^{3}}\frac{X\left(
E_{\mathbf{k}}\right)  }{4E_{\mathbf{k}}E_{\mathbf{k}+\mathbf{q}}}\nonumber\\
&  \times\left(  \frac{1}{i\Omega_{n}-E_{\mathbf{k}}-E_{\mathbf{k}+\mathbf{q}%
}}-\frac{1}{i\Omega_{n}-E_{\mathbf{k}}+E_{\mathbf{k}+\mathbf{q}}}\right.
\nonumber\\
&  \left.  +\frac{1}{i\Omega_{n}+E_{\mathbf{k}}-E_{\mathbf{k}+\mathbf{q}}%
}-\frac{1}{i\Omega_{n}+E_{\mathbf{k}}+E_{\mathbf{k}+\mathbf{q}}}\right)  .
\label{M12}%
\end{align}

Our model includes also the particular case of a Fermi superfluid near OFR,
choosing equal scattering lengths for the open and closed channels
$a_{1}=a_{2}\equiv a$ and accounting for the band offset through the chemical
potentials by $\mu_{2}=\mu_{1}-\delta/2$ with the detuning parameter $\delta$.
The scattering length $a$ and the interband coupling strength $\gamma$ are
then related to the scattering lengths for singlet and triplet channels
$\left(  a_{-},a_{+}\right)  $ by:%
\begin{align}
\frac{1}{a}  &  =\frac{1}{2}\left(  \frac{1}{a_{+}}+\frac{1}{a_{-}}\right)
,\\
\gamma &  =\frac{1}{2}\left(  \frac{1}{a_{-}}-\frac{1}{a_{+}}\right)  .
\end{align}
The set of gap equations (\ref{gapeqs}) has two solutions: the in-phase
solution with $\Delta_{2}/\Delta_{1}>0$ and the out-of phase solution with
$\Delta_{2}/\Delta_{1}<0$ \cite{Iskin3,Zhang}. For $\gamma>0$, the in-phase
solution is stable, and the out-of-phase one is metastable. It is easy to see
from (\ref{gapeqs}) that when the sign of $\gamma$ is changed ($\gamma
\rightarrow-\gamma$) at the same $1/a_{j}$, the in-phase and out-of-phase
solutions are swapped keeping the same magnitudes $\left\vert \Delta
_{j}\right\vert $ as for $\gamma>0$.

The derivation above briefly describes the analytic procedure developed in
Ref. \cite{Klimin2015} which can be found there in detail and which is
literally applied in the present work. To summarize, the GPF effective action
is the second-order term of the series for the effective pair field action
(\ref{S2B}) in powers of fluctuation coordinates. The structure of the GPF
action derived in \cite{Klimin2015} transparently reproduces known one-band
GPF actions for each band component coupled through the Josephson coupling.
For the description of collective excitations, the GPF approach is in fact
equivalent to the Random Phase Approximation (RPA) \cite{Anderson}. It is
reliable as long as fluctuation coordinates are sufficiently small, when
higher-order terms of the fluctuation expansion can be neglected.

\section{Determination of eigenmodes}

The spectrum of collective excitations can be determined using the same method
as for a one-band Fermi superfluid \cite{Kurkjian2019,Arxiv}. We solve the
equation for the determinant of the inverse fluctuation propagator
analytically continued from the set of Matsubara frequencies to the complex
frequency plane $\left(  i\Omega_{n}\rightarrow z\right)  $,%
\begin{equation}
\det\mathbb{\tilde{M}}_{2b}\left(  \mathbf{q},z\right)  =0. \label{detM}%
\end{equation}
Formally determined, this equation has no complex roots. They can be found
however following the procedure developed by P. Nozi\`{e}res \cite{Nozieres}.
A function $f\left(  z\right)  $ having a branch cut at the real axis can be
analytically continued to the lower semi-panel, $\operatorname{Im}\left(
z\right)  <0$, using the spectral function determined at the real axis
$z=\omega$,%
\begin{equation}
\rho_{f}\left(  \omega\right)  =\frac{f\left(  \omega+i0^{+}\right)  -f\left(
\omega-i0^{+}\right)  }{2\pi i}, \label{rho}%
\end{equation}
which can be analytically continued to the complex $z$ plane, $\rho_{f}\left(
\omega\right)  \rightarrow\rho_{f}\left(  z\right)  $. The analytic
continuation for $f\left(  z\right)  $ is then:%
\begin{equation}
f^{\left(  R\right)  }\left(  z\right)  =\left\{
\begin{array}
[c]{cc}%
f\left(  z\right)  , & \operatorname{Im}z>0,\\
f\left(  z\right)  -2\pi i\rho_{f}\left(  z\right)  , & \operatorname{Im}z<0.
\end{array}
\right.  \label{ancont}%
\end{equation}
The complex eigenfrequencies of collective excitations $z_{\mathbf{q}}%
\equiv\omega_{\mathbf{q}}-i\Gamma_{\mathbf{q}}/2$, accounting for both the
frequency and the damping factor, are obtained as roots of (\ref{detM}) with
the analytically continued determinant $\det\mathbb{\tilde{M}}_{2b}^{\left(
R\right)  }\left(  \mathbf{q},z\right)  $ related to $\mathbb{\tilde{M}}%
_{2b}\left(  \mathbf{q},z\right)  $ in accordance with (\ref{ancont}).

In general, the spectral function $\rho_{f}\left(  \omega\right)  $ contains
several non-analytic points which determine several windows\ for the analytic
continuation. Within the interpretation of Refs. \cite{Kurkjian2019,Arxiv},
eigenfrequencies which lie in different windows\ describe peaks of the pair
response function in the corresponding frequency intervals.

The analytic continuation method is complementary to the approach of Ref.
\cite{Zhang} where frequencies of collective modes are determined numerically
from peak positions of the density-density response function. The complex
poles of the GPF propagator give us a semianalytic solution with both
eigenfrequencies and damping factors of collective excitations, describing in
a consistent way both long-lived and damped modes. Also, the results presented
here extend the study of Leggett excitations to nonzero temperatures.

In a two-band superfluid, there exist hybrid phononic branches with sound
velocities described by EFT \cite{Klimin2015}, and two pair-breaking branches
\cite{Kurkjian2019} with frequencies strongly pinned to the two pair-breaking
continuum edges $2\Delta_{j}$. In the present treatment, we focus on Leggett
collective modes.

Further on, chemical potentials are assumed to be equal for different
spin-components in both bands: $\mu_{\uparrow,j}=\mu_{\downarrow,j}=\mu_{j}$.
First, we consider the Fermi superfluid in the BEC regime, which is typical
for Fermi gases realized using OFR. The Leggett collective modes for a Fermi
gas with OFR are studied at $T=0$ in Refs. \cite{He2016,Zhang}. Here, the
treatment is performed for nonzero temperatures. Remarkably, although the
starting fermionic model action of our work \cite{Klimin2015} is different
from that of Ref. \cite{Zhang}, they lead to the same effective bosonic actions.

We choose units such that the fermion mass $m=1/2$ and the total particle
density of two band components $n=2/\left(  3\pi^{2}\right)  $ following Ref.
\cite{Klimin2015}. In these units, the Fermi wave vector of free fermions in
two bands for zero detuning is $k_{F}\equiv\left(  3\pi^{2}n/2\right)
^{1/3}=1$, and the corresponding Fermi energy $E_{F}=1$. The equation of state
is calculated in the present work within the mean-field approximation. It is
adequate in the BCS regime, as long as a contribution of fluctuations to the
equation of state can be neglected. For stronger couplings, fluctuations
beyond the mean-field solution for the pair field lead to a significant
reduction of the superfluid transition temperature, and, consequently, to
disagreement of calculated quantities with experiment. In order to improve the
mean-field results to be more accurate predictions, we use the scaling of the
temperature dependence of Leggett mode frequencies expressing them as function
of the relative temperature $T/T_{c}$. Also the equation of state can be
written down in terms of $T/\mu_{j}$ and $\mu_{j}/\Delta_{j}$ in addition to
$\left(  T/E_{F},k_{F}a_{j}\right)  $, as in Refs. \cite{Kurkjian2019,Arxiv}.
Because fluctuations lead to scaling of parameters of state, the dimensionless
ratios $\left(  T/T_{c},T/\mu_{j},\mu_{j}/\Delta_{j}\right)  $ appear to be
independent on an equation of state, what makes the mean-field approximation
for background parameters more predictive, and results in an adequate
qualitative description of collective excitations. The scaled picture of the
eigenfrequencies can be distorted (like a conformal map) with respect to that
obtained a more precise equation of state, but all crossings and anticrossings
of different excitation branches must survive.

There are of course limitations even for this qualitative description. First,
the mean-field approximation is not justified near the superfluid transition
temperature, where there is a critical regime driven by strong fluctuations.
Thus, for quantitative agreement with experiments in the whole temperature
range below $T_{c}$, taking fluctuations into account is still necessary.
Second, anharmonic three-phonon and four-phonon processes, which are beyond
GPF, can bring a relatively important contribution to damping at low
temperatures. This analysis, as well as the detailed study of other (phononic
and pair-breaking) collective excitations in a two-band Fermi superfluid, is
beyond the scope of the present work.%

%TCIMACRO{\FRAME{ftbhFU}{3.1512in}{4.2839in}{0pt}{\Qcb{(Color online) Leggett
%mode frequencies (\QTR{em}{a}) and damping factors (\QTR{em}{b}) in the BEC
%regime as functions of detuning $\delta$ for several values of $T/T_{c}$
%calculated using the scattering lengths from Ref. \cite{Zhang}. Thin solid and
%dashed curves represent, respectively, the edges of the pair-breaking continua
%for the open-channel and closed-channel band components at $T=0$. Thin dotted
%and dot-dashed curves show the edges at $T=0.8T_{c}$.}}{\Qlb{fig:BEC}%
%}{lmx08.eps}{\special{ language "Scientific Word";  type "GRAPHIC";
%maintain-aspect-ratio TRUE;  display "ICON";  valid_file "F";
%width 3.1512in;  height 4.2839in;  depth 0pt;  original-width 5.5634in;
%original-height 7.614in;  cropleft "0";  croptop "1";  cropright "1";
%cropbottom "0";  filename 'LMX08.eps';file-properties "XNPEU";}} }%
%BeginExpansion
\begin{figure}[tbh]%
\centering
\includegraphics[
height=4.2839in,
width=3.1512in
]%
{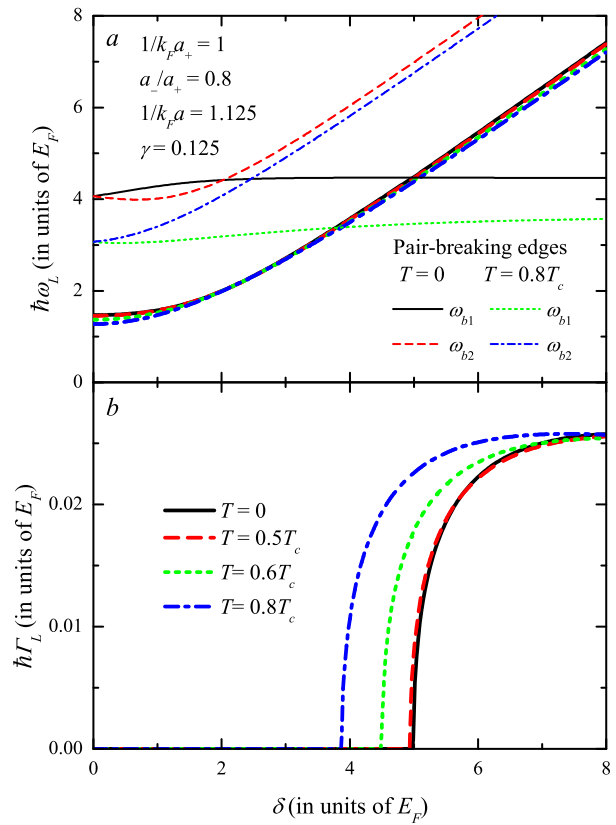}%
\caption{(Color online) Leggett mode frequencies (\emph{a}) and damping
factors (\emph{b}) in the BEC regime as functions of detuning $\delta$ for
several values of $T/T_{c}$ calculated using the scattering lengths from Ref.
\cite{Zhang}. Thin solid and dashed curves represent, respectively, the edges
of the pair-breaking continua for the open-channel and closed-channel band
components at $T=0$. Thin dotted and dot-dashed curves show the edges at
$T=0.8T_{c}$.}%
\label{fig:BEC}%
\end{figure}
%EndExpansion

In Fig. \ref{fig:BEC}, long-wavelength Leggett mode frequencies $\omega_{L}$
with the momentum $q\rightarrow0$ are plotted as a function of the detuning
factor $\delta$ for several values of $T/T_{c}$ using the same values of
scattering lengths as in Ref. \cite{Zhang}: $1/k_{F}a_{+}=1$ and $a_{-}%
/a_{+}=0.8$. This leads to the inverse scattering length $1/k_{F}a=1.\,125$
and the interband coupling $\gamma=0.125$, which correspond to $\left.
\mu\left(  T_{c}\right)  /T_{c}\right\vert _{\delta=0}\approx-0.794$ and
$\left.  \mu/\Delta\right\vert _{T=0,\delta=0}\approx-0.949$. Because
$\gamma>0$, the stable background solution for the gaps is the in-phase
solution: $\Delta_{2}/\Delta_{1}>0$. For $T=0$, the obtained Leggett mode
frequencies precisely match the results of Ref. \cite{Zhang}.

The pair-breaking continuum edge for $q=0$ is determined as a minimal value
$\omega_{b}=\min\left(  \omega_{b,1},\omega_{b,2}\right)  $ from the
pair-breaking edges for the two bands ($j=1,2$):%
\begin{equation}
\omega_{b,j}=\left\{
\begin{array}
[c]{cc}%
2\left\vert \Delta_{j}\right\vert , & \mu_{j}\geq0,\\
2\sqrt{\Delta_{j}^{2}+\mu_{j}^{2}}, & \mu_{j}<0.
\end{array}
\right.  \label{wbj}%
\end{equation}
In Fig. \ref{fig:BEC}, these two pair-breaking edges are shown by thin dashed
and dotted curves. The true pair-breaking edge $\omega_{b}$ corresponds to the
lowest of these curves. Physically, $\omega_{b}$ indicates the minimal
threshold energy above which collective excitations become damped due to decay
into fermion pairs. Within the analytic continuation method described above
(see also for details Refs. \cite{Kurkjian2019,Arxiv}), the two pair-breaking
edges determine, in general, three windows\ for the analytic continuation: the
window\ $A$ for $\omega<\omega_{b}$, the window\ $B$ for $\omega_{b}%
<\omega<\max\left(  \omega_{b,1}\omega_{b,2}\right)  $ and the window\ $C$ for
$\omega>\max\left(  \omega_{b,1}\omega_{b,2}\right)  $. The window\ $C$ does
not provide roots for Leggett modes, because it lies completely in the
pair-breaking continuum for both band components. The window\ $A$ generates
the undamped solution with the Leggett mode frequency $\omega_{L}^{\left(
A\right)  }$ which does not cross $\omega_{b}$. This undamped solution, as
shown below, exists in an interval of $\delta$ smaller than a certain critical
value $\delta_{c}$. For $\delta>\delta_{c}$, there appears a damped solution
$\omega_{L}^{\left(  B\right)  }$ provided by the window\ $B$.

The chosen values of the scattering lengths in Fig. \ref{fig:BEC} are related
to the BEC regime, where the chemical potentials for both bands are negative,
even at zero detuning. Therefore there are no pair-breaking collective
excitations in this regime \cite{Kurkjian2019}. As can be seen from the
figure, the frequencies of Leggett modes smoothly cross the pair-breaking
continuum edge $\omega_{b}$. They acquire only a relatively small damping
factor when $\omega_{L}>\omega_{b}$ but do not vanish. Therefore the visible
diminution of the Leggett mode peaks for the response function in the
continuum obtained in Ref. \cite{Zhang} can be attributed to a decrease of the
spectral weight of Leggett modes rather than to damping.%

%TCIMACRO{\FRAME{ftbhFU}{3.1862in}{2.332in}{0pt}{\Qcb{(Color online)
%\QTR{em}{Thick curves}: Leggett mode frequencies in the BEC regime as
%functions of temperature at two values of detuning $\delta=0$ and
%$\delta=4E_{F}$ calculated using the scattering lengths from Ref.
%\cite{Zhang}. \QTR{em}{Thin curves} show, respectively, the edges of the
%pair-breaking continuum for the same values of $\delta$.}}{\Qlb{fig:BECT}%
%}{lmx08t.eps}{\special{ language "Scientific Word";  type "GRAPHIC";
%maintain-aspect-ratio TRUE;  display "ICON";  valid_file "F";
%width 3.1862in;  height 2.332in;  depth 0pt;  original-width 7.6454in;
%original-height 5.5339in;  cropleft "0";  croptop "1";  cropright "1";
%cropbottom "0";  filename 'LMX08T.eps';file-properties "XNPEU";}} }%
%BeginExpansion
\begin{figure}[tbh]%
\centering
\includegraphics[
height=2.332in,
width=3.1862in
]%
{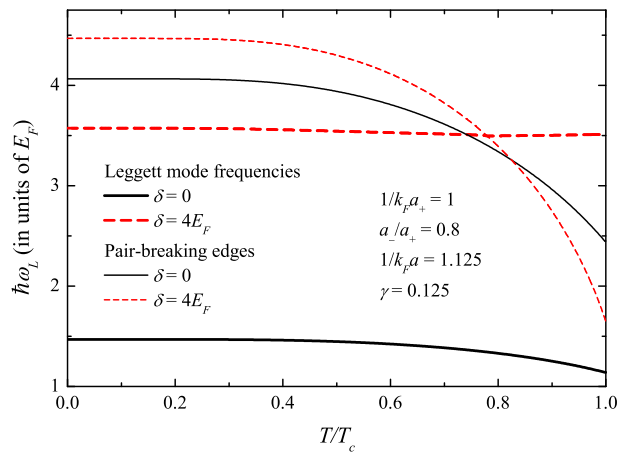}%
\caption{(Color online) \emph{Thick curves}: Leggett mode frequencies in the
BEC regime as functions of temperature at two values of detuning $\delta=0$
and $\delta=4E_{F}$ calculated using the scattering lengths from Ref.
\cite{Zhang}. \emph{Thin curves} show, respectively, the edges of the
pair-breaking continuum for the same values of $\delta$.}%
\label{fig:BECT}%
\end{figure}
%EndExpansion

In Fig. \ref{fig:BECT}, we show the temperature dependence of long-wavelength
Leggett mode frequencies and pair-breaking continuum edges for a Fermi
superfluid using the same parameters as in Fig. \ref{fig:BEC}, with two values
of detuning $\delta=0$ and $\delta=4E_{F}$. As distinct from the BCS/unitarity
cases, where the Leggett mode frequencies tend to zero at $T\rightarrow T_{c}%
$, in the BEC regime they remain finite when approaching the transition
temperature. This difference occurs due to the following reasons. First,
previous theoretical studies of Leggett collective excitations were based on a
perturbative weak-coupling approach, which is focused on the BCS regime but
can fail at strong coupling. In the weak-coupling approach, $\omega_{L}%
^{2}\propto\Delta_{1}\Delta_{2}$ and therefore the Leggett mode frequency
turns to zero at $T_{c}$. The present treatment does not assume the
weak-coupling approximation, and therefore the aforesaid trend is not
necessarily fulfilled. Moreover, in the BEC regime, as soon as $\mu<0$, the
pair-breaking continuum edge exceeds the value $2\Delta_{2}$ according to
(\ref{wbj}), and hence $\omega_{b,j}\neq0$ at the transition temperature,
which favors the survival of Leggett modes. As can be seen from Figs.
\ref{fig:BEC} (\emph{b}) and \ref{fig:BECT}, the Leggett mode frequencies
cross the pair-breaking continuum edge only at a sufficiently large detuning
$\delta_{c}$. The value $\delta_{c}$ slightly decreases when increasing
temperature [see Fig. \ref{fig:BEC} (\emph{b})] but it does not turn to zero.
This behavior is specific for the sufficiently far BEC regime, but this not
always the case for weaker couplings, where the Leggett mode frequency may
reach $\omega_{b}$, as shown below. We can conclude that the BEC regime is
promising for detection of Leggett collective excitations.%

%TCIMACRO{\FRAME{ftbhFU}{3.0294in}{4.1843in}{0pt}{\Qcb{(Color online)
%\QTR{em}{Solid curves}: Leggett mode frequencies $\omega_{L}^{\left(
%A\right)  },\omega_{L}^{\left(  B\right)  }$. \QTR{em}{Dashed curves}: the
%damping factor of the second root of the dispersion equation. \QTR{em}{Dotted
%curves}: the Leggett mode frequency given by the low-frequency expansion of
%the inverse GPF propagator. \QTR{em}{The dot-dashed curves} show $2\Delta_{2}$
%as a function of temperature. The vertical line indicates the temperature at
%which $\omega_{L}^{\left(  B\right)  }$ crosses the pair-breaking continuum
%edge $2\Delta_{2}$. The calculation is performed for $1/a_{1}=0$,
%$1/a_{2}=-0.5$, $\gamma=0.02$ (\QTR{em}{a}) and $\gamma=0.1$ (\QTR{em}{b}).}}%
%{\Qlb{fig:LM1}}{lmcross.eps}{\special{ language "Scientific Word";
%type "GRAPHIC";  maintain-aspect-ratio TRUE;  display "ICON";
%valid_file "F";  width 3.0294in;  height 4.1843in;  depth 0pt;
%original-width 5.4592in;  original-height 7.5578in;  cropleft "0";
%filename 'LMcross.eps';file-properties "XNPEU";}} }%
%BeginExpansion
\begin{figure}[tbh]%
\centering
\includegraphics[
height=4.1843in,
width=3.0294in
]%
{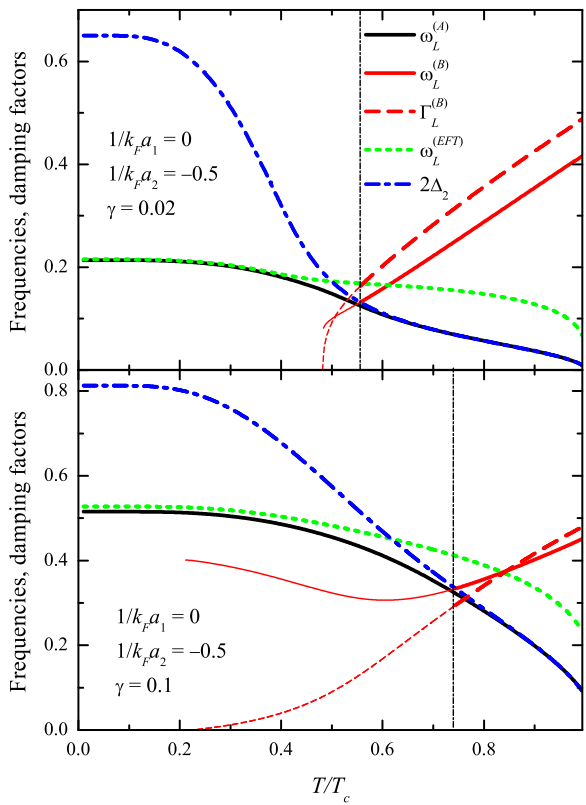}%
\caption{(Color online) \emph{Solid curves}: Leggett mode frequencies
$\omega_{L}^{\left(  A\right)  },\omega_{L}^{\left(  B\right)  }$.
\emph{Dashed curves}: the damping factor of the second root of the dispersion
equation. \emph{Dotted curves}: the Leggett mode frequency given by the
low-frequency expansion of the inverse GPF propagator. \emph{The dot-dashed
curves} show $2\Delta_{2}$ as a function of temperature. The vertical line
indicates the temperature at which $\omega_{L}^{\left(  B\right)  }$ crosses
the pair-breaking continuum edge $2\Delta_{2}$. The calculation is performed
for $1/a_{1}=0$, $1/a_{2}=-0.5$, $\gamma=0.02$ (\emph{a}) and $\gamma=0.1$
(\emph{b}).}%
\label{fig:LM1}%
\end{figure}
%EndExpansion

We consider also Leggett modes far from the BEC regime, and with different
scattering lengths for open and closed channels, which is more general than
the case of OFR where $a_{1}=a_{2}$. This setup is not yet reached
experimentally, but can represent an interest for future experiments. Fig.
\ref{fig:LM1} shows Leggett mode frequencies at $q=0$ as a function of
temperature for the interband coupling $\gamma=0.02$ (a) and $\gamma=0.1$ (b).
The inverse scattering lengths are taken here $1/a_{1}=0$ and $1/a_{2}=-0.5$.
These parameters give us the ratios $\left.  \mu\left(  T_{c}\right)
/T_{c}\right\vert _{\delta=0}\approx1.50$, $\left.  \mu/\Delta_{1}\right\vert
_{T=0,\delta=0}\approx0.847$, $\left.  \mu/\Delta_{2}\right\vert
_{T=0,\delta=0}\approx2.11$ for the interband coupling $\gamma=0.02$ (panel
\emph{a}) and $\left.  \mu\left(  T_{c}\right)  /T_{c}\right\vert _{\delta
=0}\approx1.38$, $\left.  \mu/\Delta_{1}\right\vert _{T=0,\delta=0}%
\approx0.777$, $\left.  \mu/\Delta_{2}\right\vert _{T=0,\delta=0}\approx1.59$
for $\gamma=0.1$ (panel \emph{b}). The obtained solutions are compared with
the Leggett mode frequency $\omega_{L}^{\left(  EFT\right)  }$ obtained using
the low-frequency expansion of the effective action \cite{Klimin2015} and with
the energy of the pair-breaking continuum edge $2\Delta_{2}$.

At low temperatures $T\ll T_{c}$, only one root of the dispersion equation
exists, with the frequency denoted here as $\omega_{L}^{\left(  A\right)
}<2\Delta_{2}$ and shown as a black curve in Fig. \ref{fig:LM1}. For $q=0$,
the Leggett mode corresponding to this root is undamped, because the
quasiparticle-quasiparticle and quasihole-quasihole terms [with $i\Omega
_{n}\pm\left(  E_{\mathbf{k}}+E_{\mathbf{k}+\mathbf{q}}\right)  $] in the GPF
matrix elements (\ref{M11}) and (\ref{M12}) cannot contribute to the damping
of collective excitations with energy below $2\Delta_{2}$, while
quasiparticle-quasihole\ terms [with $i\Omega_{n}\pm\left(  E_{\mathbf{k}%
}-E_{\mathbf{k}+\mathbf{q}}\right)  $] do not contribute at all to the matrix
elements when $q=0$. This picture is quite similar to that of pair-breaking
collective modes \cite{Kurkjian2019}.

As the temperature is increased, the Leggett mode frequency $\omega
_{L}^{\left(  A\right)  }$ at $q=0$ tends to $2\Delta_{2}$, remaining
undamped. It does not cross the pair-breaking continuum edge. The underlying
physics of this behavior consists in the avoided crossing with the
pair-breaking collective excitations, because different eigenmodes have
different energies. This anticrossing exists when the chemical potential
$\mu_{2}>0$, particularly in the BCS/unitarity regime. It does not exist in
the BEC case considered above because of the absence of pair-breaking
collective excitations.

The obtained anticrossing of the Leggett mode frequency is drastically
different from the solution $\omega_{L}^{\left(  EFT\right)  }$ predicted by
the low-energy expansion shown as a green curve in Fig. \ref{fig:LM1}. The
latter one does not capture the interplay of the Leggett collective mode
with\ the pair-breaking continuum edge and hence crosses the value
$\omega=2\Delta_{2}$ without any feature.

At sufficiently high temperatures, a second eigenfrequency $\omega
_{L}^{\left(  B\right)  }$ appears, which is strongly damped in the considered
regime far from BEC. The frequency and the damping factor for this root are
shown in Fig. \ref{fig:LM1}, respectively, by solid and dashed red curves. In
order to interpret it properly, we recall the analytic continuation of the
matrix elements through the branch cut. For $q=0$, there are 4 non-analytic
points on the real axis: $\omega_{1}=2\Delta_{2}$, $\omega_{2}=2\sqrt
{\Delta_{2}^{2}+\mu^{2}}$, $\omega_{3}=2\Delta_{1}$, $\omega_{4}=2\sqrt
{\Delta_{1}^{2}+\mu^{2}}$. The solution $\omega_{L}^{\left(  B\right)
}-i\Gamma_{L}^{\left(  B\right)  }/2$ appears when the analytic continuation
is performed through the window\ $B$,\ $\omega_{1}<\omega<\min\left(
\omega_{2},\omega_{3}\right)  $. Strictly speaking, this solution is
physically relevant (e. g., for the spectral response) only in the
window\ $B$, where its frequency and damping are shown by thick red curves in
Fig. \ref{fig:LM1}. We can formally extend the analytic continuation through
the window $\omega_{1}<\omega<\min\left(  \omega_{2},\omega_{3}\right)  $ to
the whole lower half-plane, as in Ref. \cite{Kurkjian2019}: this solution is
shown by thin curves. We can see that the second root appears at lower
temperatures starting from a finite $\omega_{L}^{\left(  B\right)  }$ with
zero damping. As temperature rises, the damping factor of this mode rapidly increases.

We can see that the full solution of the reduced dispersion equation
corresponding to Leggett modes can be close to the result of the low-frequency
expansion $\omega_{L}^{\left(  EFT\right)  }$ only when the temperature is
sufficiently low. The low-frequency expansion thus becomes inapplicable when
the Leggett mode frequency approaches the range close to the pair-breaking
continuum edge.

It is a complicated question whether the damped eigenfrequency $\omega
_{L}^{\left(  B\right)  }-i\Gamma_{L}^{\left(  B\right)  }/2$ can be
attributed to the pair-breaking collective excitation, because the latter one
in a one-band system reveals at small $q$ strong pinning to the pair-breaking
continuum edge. Also it hardly can be a purely phase Leggett collective mode,
because amplitude fluctuations must bring a substantial contribution to
collective excitations in the continuum. In general, this solution is a hybrid
of phase and amplitude fluctuations. In order to clarify the physical
interpretation of the obtained solutions, it is useful to consider also the
spectral weight functions for the fluctuation propagator at a nonzero
momentum, since the pair-breaking modes disappear at $q=0$.

\section{Collective excitations for a nonzero momentum in spectral weight
functions}

In the literature, phononic and Leggett modes are associated with phase
collective excitations, while pair-breaking modes are attributed to amplitude
excitations. This is correct only in the far BCS limit, where the phase and
amplitude modes can be accurately identified through the poles of the
fluctuation propagator. In the general case, amplitude and phase oscillations
are coupled, and one cannot extract them explicitly. Thus both amplitude and
phase excitations bring contributions to all modes, and we can see
fingerprints of all modes in all response functions. The dominating terms
remain however the same as in the BCS case.

\begin{figure}[tbh]%
\centering
\includegraphics[
height=3.3218in,
width=3.2886in
]%
{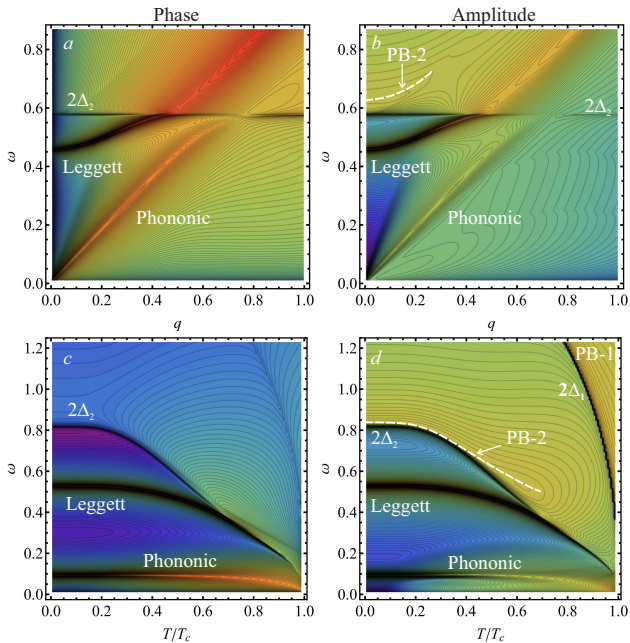}%
\caption{(Color online) (\emph{a,~b}) Contour plots of the phase-phase and
amplitude-amplitude spectral weight functions for a two-band Fermi gas with
$1/k_{F}a_{1}=0,1/k_{F}a_{2}=-0.5$ and $\gamma=0.1$ in the variables $\left(
q,\omega\right)  $ at $T=0.5T_{c}$. (\emph{c,~d}) Spectral weight functions in
the variables $\left(  T,\omega\right)  $ for $q=0.1k_{F}$. Dashed curves in
panels (\emph{b,~d}) are guides for the eye indicating pair-breaking modes for
the \textquotedblleft weak\textquotedblright\ band component.}%
\label{ContourBCS}%
\end{figure}

\begin{figure}[tbh]%
\centering
\includegraphics[
height=3.414in,
width=3.2065in
]%
{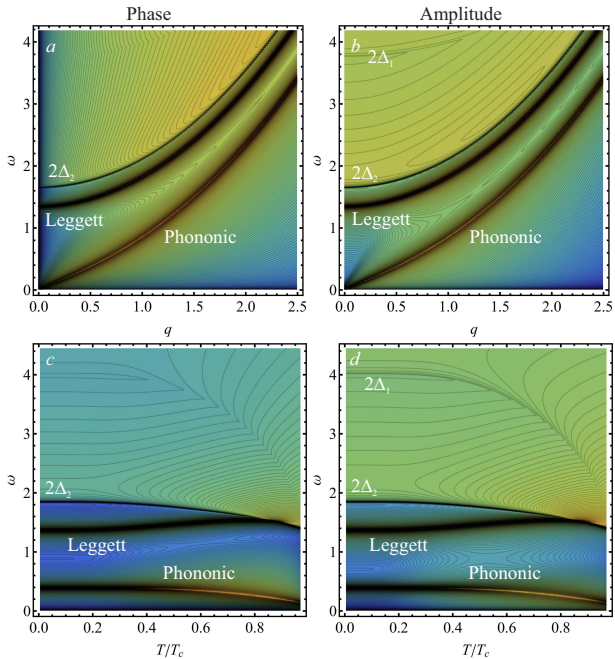}%
\caption{(Color online) (\emph{a,~b}) Contour plots of the phase-phase and
amplitude-amplitude spectral weight functions for a two-band Fermi gas with
$1/k_{F}a_{1}=1,1/k_{F}a_{2}=0.5$ and $\gamma=0.1$ in the variables $\left(
q,\omega\right)  $ at $T=0.5T_{c}$. (\emph{c,~d}) Spectral weight functions in
the variables $\left(  T,\omega\right)  $ for $q=0.5k_{F}$.}%
\label{ContourBEC}%
\end{figure}

Consequently, also an alternative method can be used to determine spectra of
collective excitations in a superfluid Fermi gas, namely through the spectral
weight functions for the GPF propagator \cite{Han}, which reveals the presence
of poles via peaks. The expressions for the spectral weight functions are
determined in Appendix A. Here, we discuss relative phase-phase and
amplitude-amplitude spectral weight functions, because they are the most
representative to extract necessary physical information on collective
excitations. The spectral functions (multiplied by $q^{2}$) are shown as
contour plots in Figs. \ref{ContourBCS} and \ref{ContourBEC}.%

%TCIMACRO{\FRAME{ftbhFU}{3.2886in}{3.3218in}{0pt}{\Qcb{(Color online)
%(\QTR{em}{a,~b}) Contour plots of the phase-phase and amplitude-amplitude
%spectral weight functions for a two-band Fermi gas with $1/k_{F}%
%a_{1}=0,1/k_{F}a_{2}=-0.5$ and $\gamma=0.1$ in the variables $\left(
%q,\omega\right)  $ at $T=0.5T_{c}$. (\QTR{em}{c,~d}) Spectral weight functions
%in the variables $\left(  T,\omega\right)  $ for $q=0.1k_{F}$. Dashed curves
%in panels (\QTR{em}{b,~d}) are guides for the eye indicating pair-breaking
%modes for the \textquotedblleft weak\textquotedblright\ band component.}}%
%{\Qlb{ContourBCS}}{lmbcs.eps}{\special{ language "Scientific Word";
%type "GRAPHIC";  maintain-aspect-ratio TRUE;  display "ICON";
%valid_file "F";  width 3.2886in;  height 3.3218in;  depth 0pt;
%original-width 7.5845in;  original-height 7.662in;  cropleft "0";
%filename 'LMBCS.eps';file-properties "XNPEU";}} }%
%BeginExpansion
%EndExpansion

Fig. \ref{ContourBCS} describes the momentum and temperature behavior of
collective excitations in the unitarity/BCS regime, with the inverse
scattering lengths $1/k_{F}a_{1}=0$, $1/k_{F}a_{2}=-0.5$ and the interband
coupling $\gamma=0.1$. Since the Leggett modes below the pair-breaking
continuum edge (i.~e., with $\omega<\omega_{b,2}$) have a very low damping
even at nonzero momentum, the spectral weight functions are plotted with the
complex argument $z=\omega+i\gamma$ where a relatively small damping parameter
$\gamma=0.03E_{F}$ is added in order to visualize Leggett modes. The spectra
of collective excitations are shown as functions of momentum and temperature.
The phase-phase spectral weight functions in Figs. \ref{ContourBCS}%
~(\emph{a},~\emph{c}) exhibit two branches of collective excitations: Leggett
and phononic modes. These branches are also visible in Figs. \ref{ContourBCS}%
~(\emph{b},~\emph{d}), which show amplitude-amplitude spectral weight
functions, but with smaller magnitudes. This confirms that these branches are
dominated by phase fluctuations. This is especially expressed for phononic
excitations, whereas Leggett modes in the unitarity/BCS regime contain a
non-negligible part of amplitude fluctuations, as distinct from the far BCS
limit, where they are purely phase excitations.

The pair-breaking excitations are not manifested in the phase spectral
weights, since they are essentially governed by amplitude fluctuations. They
are revealed as finite-width peaks denoted as PB$_{1}$ and PB$_{2}$ in Figs.
\ref{ContourBCS}~(\emph{b},~\emph{d}). The pair-breaking excitation branch for
the \textquotedblleft weak\textquotedblright\ (with $1/k_{F}a_{2}=-0.5$)\ band
PB$_{2}$ is clearly visible only for sufficiently low temperatures, here at
$T\lesssim0.5T_{c}$. When temperature rises, this pair-breaking excitation
peak shifts to higher frequencies away from the pair-breaking continuum edge
and gradually broadens. As can be seen from Fig. \ref{ContourBCS}~(\emph{d}),
this shift shows an avoided crossing with Leggett modes, whose frequencies
tend to $2\Delta_{2}$ from below. This behavior of pair-breaking and Leggett
modes makes clear the interpretation of the finite-width peak above
$2\Delta_{2}$ in Fig. \ref{fig:LM1}: we can attribute it to damped
pair-breaking modes, unpinned\ from $2\Delta_{2}$ due to anticrossing with the
Leggett branch. Moreover, due to this anticrossing, pair-breaking modes in a
two-band Fermi gas become visible even in the zero-momentum limit, as distinct
from the one-band system.

At large momenta, the Leggett branch of collective excitations in Fig.
\ref{ContourBCS}~(\emph{a}) is continued through the pair-breaking continuum
edge without anticrossing with the pair-breaking branch, since the latter one
dissolves at sufficiently large $q$. Instead of this, the Leggett excitation
branch exhibits anticrossing with the phononic branch when passing the edge.
As one can see from the comparison of Figs. \ref{ContourBCS}~(\emph{a}) and
(\emph{c}), it has more expressed phase character above $2\Delta_{2}$, because
the amplitude spectral weight for this branch diminishes. We can conclude
therefore that the Leggett branch above the pair-breaking continuum edge
acquires a significant contribution of phase fluctuations coming from the
phononic branch, and becomes a mix of phononic and Leggett modes.

In Figs \ref{ContourBCS}~(\emph{c,~d}), we can observe also the upper
pair-breaking frequency $\omega_{b,1}=2\Delta_{1}$. In agreement with the
above results for the zero-momentum limit, there exist a pair-breaking
collective excitation branch close to $\omega_{b,1}$, which is absent in the
phase-phase spectral weight function, Fig. \ref{ContourBCS}~(\emph{c}), and is
clearly seen in the amplitude-amplitude spectrum, Fig. \ref{ContourBCS}%
~(\emph{d}).%

%TCIMACRO{\FRAME{ftbhFU}{3.2065in}{3.414in}{0pt}{\Qcb{(Color online)
%(\QTR{em}{a,~b}) Contour plots of the phase-phase and amplitude-amplitude
%spectral weight functions for a two-band Fermi gas with $1/k_{F}%
%a_{1}=1,1/k_{F}a_{2}=0.5$ and $\gamma=0.1$ in the variables $\left(
%q,\omega\right)  $ at $T=0.5T_{c}$. (\QTR{em}{c,~d}) Spectral weight functions
%in the variables $\left(  T,\omega\right)  $ for $q=0.5k_{F}$.}}%
%{\Qlb{ContourBEC}}{lmbec.eps}{\special{ language "Scientific Word";
%type "GRAPHIC";  maintain-aspect-ratio TRUE;  display "ICON";
%valid_file "F";  width 3.2065in;  height 3.414in;  depth 0pt;
%original-width 7.5688in;  original-height 8.0633in;  cropleft "0";
%filename 'LMBEC.eps';file-properties "XNPEU";}} }%
%BeginExpansion
%EndExpansion

The spectral weight functions for a two-band system in the BEC regime with
$1/k_{F}a_{1}=1$, $1/k_{F}a_{2}=0.5$ and $\gamma=0.1$ plotted in Fig.
\ref{ContourBEC} show a different behavior of the collective excitations. In
the BEC regime the pair-breaking collective excitations do not exist.
\cite{Kurkjian2019}, hence there are no peaks corresponding to these branches
in the spectral weight functions. As can be seen from Figs. \ref{ContourBEC}%
~(\emph{c,~d}), the Leggett branch in the BEC regime can approach the
pair-breaking continuum edge $2\Delta_{2}$ only at a temperature rather close
to $T_{c}$. Since pair-breaking excitations are absent in the BEC regime, the
Leggett branch can cross the pair-breaking continuum edge, becoming
damped,which can also be observed in Figs. \ref{ContourBEC}~(\emph{c,~d}). For
lower temperatures, e. g., $T=0.5T_{c}$ as shown in Figs. \ref{ContourBEC}%
~(\emph{a,~b}), both phononic and Leggett modes vary similarly to each other
and to $2\Delta_{2}$ without crossing. This behavior is simpler than in the
unitarity/BCS regimes, but represents an interest, because the BEC regime has
been realized in experiments \cite{Pagano,Hofer}. As can be seen from Fig.
\ref{ContourBEC}, Leggett and phononic collective excitations in the BEC
regime have much stronger contributions from amplitude fluctuations than in
the BCS regime.

\section{Conclusions}

In our preceding works \cite{Kurkjian2019,Arxiv}, we have determined for a
one-band system the frequency and the damping factor for phononic and
pair-breaking collective excitations in a nonperturbative way using the
analytic continuation for the Gaussian fluctuation propagator. Here, this
method has been straightforwardly extended for two-band Fermi superfluids,
which reveal Leggett collective excitations absent in one-band systems.

The behavior of the phase collective excitations for interacting two-band
Fermi gases was already studied in the limiting cases $T=0$ and $T\rightarrow
T_{c}$ \cite{Iskin1,Iskin2,Zhang}. They mix with each other, resulting in one
phononic and one Leggett branch, common for the whole two-band system. The
question of existence of the two pair-breaking amplitude branches in a
two-band Fermi gas is more subtle. In a one-band Fermi gas with $s$-wave
pairing, the GPF approximation reveals up to two branches of collective
excitations: the phononic modes \cite{Arxiv}, and except the BEC regime, the
pair-breaking modes \cite{Kurkjian2019}. Intuitively, it could be expected
that the number of branches would be conserved when the two band components
interact. It was not however \emph{a priori} clear whether the second
pair-breaking mode for a weak\ band survives. We have found that the GPF
approximation for the two-band Fermi gas predicts up to four branches of
collective excitations. In the unitarity/BCS regimes, there exist one phononic
branch, one Leggett branch and two pair-breaking branches in a two-band Fermi
gas. Phononic and Leggett modes consist mainly of phase fluctuations, and
pair-breaking modes are related to amplitude fluctuations. The first two
branches \emph{represent a strong mix} of the fluctuations for both band
components, while the pair-breaking modes of the two band components are
essentially pinned to the two pair-breaking edges. They therefore couple only
weakly to each other. However, the pair-breaking collective excitations can
strongly interact with Leggett modes, revealing an avoided crossing near the
pair-breaking continuum edge. Moreover, the interaction with Leggett
excitations reveals the pair-breaking modes even in the long-wavelength limit,
where they vanish in a one-band system.

We have found that in the BEC regime for a Fermi superfluid prepared using
OFR, the Leggett mode as a function of detuning does not vanish when merging
with the pair-breaking continuum, but becomes damped. There is a drastic
difference between the behavior of Leggett collective excitations near the
pair-breaking continuum edge in the BEC regime and in the crossover regime
other than BEC. In the BEC regime, the Leggett mode passes the edge without
any feature. Moreover, within the present nonperturbative treatment, Leggett
mode frequencies in the BEC regime do not turn to zero at the transition
temperature, as distinct from weak-coupling results. On the contrary, far from
BEC, the Leggett mode frequency avoids crossing with the edge. This different
behavior can be attributed to the interplay of Leggett and pair-breaking
collective excitations, which exist in the BCS-BEC crossover sufficiently
close to BCS, but disappear in the BEC regime. We can conclude that
strong-coupling two-band Fermi superfluids, particularly near the orbital
Feshbach resonance, are favorable for the experimental observation of Leggett
collective excitations. At present, only the BEC case has been experimentally
achieved for two-band atomic gases, but coupling parameters of a Fermi gas
close to OFR can be experimentally tuned through the whole BCS-BEC crossover
region and even into the BCS regime \cite{Deng}. This makes the present study
relevant for subsequent experiments.

Collective modes of the pair field can be visible through the density response
due to the coupling between pair and density fields
\cite{Zhang,Arxiv,Castin2001}. An experimental observation of Leggett
collective excitations is a challenging problem, because they, in general,
have relatively small spectral weights or high decay rates. Leggett modes have
been detected in a two-band superconductor MgB$_{2}$ using Raman
\cite{Blumberg} and tunneling \cite{Ponomarev} spectroscopy, and, recently,
using intense terahertz lights pulses \cite{Giorgianni}. In the experimentally
realized two-band $^{173}$Yb cold atomic gases with OFR, Leggett collective
excitations can be hardly resolved, because they are strongly damped in the
experimentally achieved stable out-of-phase state \cite{He2016}. For two-band
Fermi gases with both large singlet and triplet scattering lengths, the whole
BCS-BEC crossover, which is favorable for long-sought Leggett modes, is
promising for future experiments.

The GPF formalism exploited in the present work is based on the model action
functional, which is suitable for charge-neutral Fermi gases in the BCS-BEC
crossover rather than for superconductors. It is possible to reformulate the
method using a BCS-like model, so that it will allow us to calculate in the
same way both frequencies and damping rates of collective excitations in
multiband superconductors, as MgB$_{2}$ and iron-based multiband materials.

\begin{acknowledgments}
We are grateful to C. A. R. S\'{a} de Melo for valuable discussions. This
research was supported by the University Research Fund (BOF) of the University
of Antwerp and by the Flemish Research Foundation (FWO-Vl), project No.
G.0429.15.N. and the European Union's Horizon 2020 research and innovation
program under the Marie Sk\l odowska-Curie grant agreement number 665501.
\end{acknowledgments}

%

%TCIMACRO{\TeXButton{Appendix}{\appendix}}%
%BeginExpansion
\appendix
%EndExpansion

\section{Spectral weight functions for a two-band system}

The spectral weight functions for the GPF effective action $S_{GPF}^{\left(
2b\right)  }$ are determined using the generating functional $P\left[
\bar{\eta},\eta\right]  $ which depends on the auxiliary field variables:%
\begin{equation}
P\left[  \bar{\eta},\eta\right]  =\left\langle e^{-\sum_{\mathbf{q},n}\left(
\bar{\eta}_{\mathbf{q},n}\varphi_{\mathbf{q},n}+\bar{\varphi}_{\mathbf{q}%
,n}\eta_{\mathbf{q},n}\right)  }\right\rangle _{S_{GPF}^{\left(  2b\right)  }%
}, \label{gen}%
\end{equation}
where, following the notations of Eq.(\ref{Nambu}), $\eta_{\mathbf{q},n}%
,\bar{\eta}_{\mathbf{q},n}$ are 4-dimensional vectors, similarly to
$\varphi_{\mathbf{q},n},\bar{\varphi}_{\mathbf{q},n}$. This generating
functional is calculated analytically using a linear shift of the pair field
variables, which results in the expression:%
\begin{equation}
P\left[  \bar{\eta},\eta\right]  =\exp\left(  \frac{1}{2}\sum_{\mathbf{q}%
,n}\bar{\eta}_{\mathbf{q},n}\mathbb{M}_{2b}^{-1}\left(  \mathbf{q},i\Omega
_{n}\right)  \eta_{\mathbf{q},n}\right)  . \label{gen2}%
\end{equation}
The spectral weight functions in the Matsubara representation are then
determined by the matrix elements of the GPF propagator, which is the inverse
of $\mathbb{M}_{2b}$:%
\begin{equation}
\left\langle \left(  \bar{\varphi}_{\mathbf{q},n}\right)  _{j}\left(
\varphi_{\mathbf{q},n}\right)  _{l}\right\rangle =\left[  \mathbb{M}_{2b}%
^{-1}\left(  \mathbf{q},i\Omega_{n}\right)  \right]  _{j,l}. \label{avs}%
\end{equation}
The matrix elements of the GPF propagator are explicitly given by:
%TCIMACRO{\TeXButton{Begin widetext}{\begin{widetext}} }%
%BeginExpansion
\begin{widetext}
%EndExpansion
\begin{equation}
\mathbb{M}_{2b}^{-1}=\left(
\begin{array}
[c]{cccc}%
\frac{\tilde{M}_{22}^{\left(  1\right)  }\det\mathbb{\tilde{M}}^{\left(
2\right)  }-\varkappa^{2}\tilde{M}_{11}^{\left(  2\right)  }}{\det
\mathbb{M}_{2b}} & -\frac{\tilde{M}_{12}^{\left(  1\right)  }\det
\mathbb{\tilde{M}}^{\left(  2\right)  }+\varkappa^{2}\tilde{M}_{12}^{\left(
2\right)  }}{\det\mathbb{M}_{2b}} & \varkappa\frac{\tilde{M}_{12}^{\left(
1\right)  }\tilde{M}_{12}^{\left(  2\right)  }+\tilde{M}_{22}^{\left(
1\right)  }\tilde{M}_{22}^{\left(  2\right)  }-\varkappa^{2}}{\det
\mathbb{M}_{2b}} & -\varkappa\frac{\tilde{M}_{12}^{\left(  1\right)  }%
\tilde{M}_{11}^{\left(  2\right)  }+\tilde{M}_{22}^{\left(  1\right)  }%
\tilde{M}_{12}^{\left(  2\right)  }}{\det\mathbb{M}_{2b}}\\
-\frac{\tilde{M}_{12}^{\left(  1\right)  }\det\mathbb{\tilde{M}}^{\left(
2\right)  }+\varkappa^{2}\tilde{M}_{12}^{\left(  2\right)  }}{\det
\mathbb{M}_{2b}} & \frac{\tilde{M}_{11}^{\left(  1\right)  }\det
\mathbb{\tilde{M}}^{\left(  2\right)  }-\varkappa^{2}\tilde{M}_{22}^{\left(
2\right)  }}{\det\mathbb{M}_{2b}} & -\varkappa\frac{\tilde{M}_{11}^{\left(
1\right)  }\tilde{M}_{12}^{\left(  2\right)  }+\tilde{M}_{12}^{\left(
1\right)  }\tilde{M}_{22}^{\left(  2\right)  }}{\det\mathbb{M}_{2b}} &
\varkappa\frac{\tilde{M}_{11}^{\left(  1\right)  }\tilde{M}_{11}^{\left(
2\right)  }+\tilde{M}_{12}^{\left(  1\right)  }\tilde{M}_{12}^{\left(
2\right)  }-\varkappa^{2}}{\det\mathbb{M}_{2b}}\\
\varkappa\frac{\tilde{M}_{12}^{\left(  1\right)  }\tilde{M}_{12}^{\left(
2\right)  }+\tilde{M}_{22}^{\left(  1\right)  }\tilde{M}_{22}^{\left(
2\right)  }-\varkappa^{2}}{\det\mathbb{M}_{2b}} & -\varkappa\frac{\tilde
{M}_{11}^{\left(  1\right)  }\tilde{M}_{12}^{\left(  2\right)  }+\tilde
{M}_{12}^{\left(  1\right)  }\tilde{M}_{22}^{\left(  2\right)  }}%
{\det\mathbb{M}_{2b}} & \frac{\tilde{M}_{22}^{\left(  2\right)  }%
\det\mathbb{\tilde{M}}^{\left(  1\right)  }-\tilde{M}_{11}^{\left(  1\right)
}\varkappa^{2}}{\det\mathbb{M}_{2b}} & -\frac{\tilde{M}_{12}^{\left(
2\right)  }\det\mathbb{\tilde{M}}^{\left(  1\right)  }+\tilde{M}_{12}^{\left(
1\right)  }\varkappa^{2}}{\det\mathbb{M}_{2b}}\\
-\varkappa\frac{\tilde{M}_{12}^{\left(  1\right)  }\tilde{M}_{11}^{\left(
2\right)  }+\tilde{M}_{22}^{\left(  1\right)  }\tilde{M}_{12}^{\left(
2\right)  }}{\det\mathbb{M}_{2b}} & \varkappa\frac{\tilde{M}_{11}^{\left(
1\right)  }\tilde{M}_{11}^{\left(  2\right)  }+\tilde{M}_{12}^{\left(
1\right)  }\tilde{M}_{12}^{\left(  2\right)  }-\varkappa^{2}}{\det
\mathbb{M}_{2b}} & -\frac{\tilde{M}_{12}^{\left(  2\right)  }\det
\mathbb{\tilde{M}}^{\left(  1\right)  }+\tilde{M}_{12}^{\left(  1\right)
}\varkappa^{2}}{\det\mathbb{M}_{2b}} & \frac{\tilde{M}_{11}^{\left(  2\right)
}\det\mathbb{\tilde{M}}^{\left(  1\right)  }-\varkappa^{2}\tilde{M}%
_{22}^{\left(  1\right)  }}{\det\mathbb{M}_{2b}}%
\end{array}
\right)  . \label{inv}%
\end{equation}%
%TCIMACRO{\TeXButton{End widetext}{\end{widetext}}}%
%BeginExpansion
\end{widetext}%
%EndExpansion
This matrix gives us, in particular, the intraband and interband spectral
weight functions:%
\begin{align}
\left\langle \bar{\varphi}_{\mathbf{q},n}^{\left(  1\right)  }\varphi
_{\mathbf{q},n}^{\left(  1\right)  }\right\rangle  &  =\left[  \mathbb{M}%
_{2b}^{-1}\left(  \mathbf{q},i\Omega_{n}\right)  \right]  _{11},\label{r1}\\
\left\langle \bar{\varphi}_{\mathbf{q},n}^{\left(  2\right)  }\varphi
_{\mathbf{q},n}^{\left(  2\right)  }\right\rangle  &  =\left[  \mathbb{M}%
_{2b}^{-1}\left(  \mathbf{q},i\Omega_{n}\right)  \right]  _{33},\label{r2}\\
\left\langle \bar{\varphi}_{\mathbf{q},n}^{\left(  1\right)  }\varphi
_{\mathbf{q},n}^{\left(  2\right)  }\right\rangle  &  =\left[  \mathbb{M}%
_{2b}^{-1}\left(  \mathbf{q},i\Omega_{n}\right)  \right]  _{13},\label{r3}\\
\left\langle \bar{\varphi}_{\mathbf{q},n}^{\left(  2\right)  }\varphi
_{\mathbf{q},n}^{\left(  1\right)  }\right\rangle  &  =\left[  \mathbb{M}%
_{2b}^{-1}\left(  \mathbf{q},i\Omega_{n}\right)  \right]  _{31}. \label{r4}%
\end{align}

Analytically continuing the Matsubara frequencies to the complex $z$ plane, we
determine the spectral weight functions for the total and relative responses,%
\begin{align}
&  \chi^{\left(  \pm\right)  }\left(  \mathbf{q},z\right) \nonumber\\
&  =\frac{1}{\pi}\operatorname{Im}\left.  \left\langle \left(  \bar{\varphi
}_{\mathbf{q},n}^{\left(  1\right)  }\pm\bar{\varphi}_{\mathbf{q},n}^{\left(
2\right)  }\right)  \left(  \varphi_{\mathbf{q},n}^{\left(  1\right)  }%
\pm\varphi_{\mathbf{q},n}^{\left(  2\right)  }\right)  \right\rangle
\right\vert _{i\Omega_{n}\rightarrow z}, \label{hi1}%
\end{align}
which are expressed through the matrix elements of the GPF propagator using
(\ref{r1}) to (\ref{r4}):%
\begin{align}
\chi^{\left(  \pm\right)  }\left(  \mathbf{q},\omega\right)   &  =\frac{1}%
{\pi}\operatorname{Im}\left\{  \left[  \mathbb{M}_{2b}^{-1}\left(
\mathbf{q},\omega+i0^{+}\right)  \right]  _{22}\right. \nonumber\\
&  +\left[  \mathbb{M}_{2b}^{-1}\left(  \mathbf{q},\omega+i0^{+}\right)
\right]  _{44}\nonumber\\
&  \left.  \pm2\left[  \mathbb{M}_{2b}^{-1}\left(  \mathbf{q},\omega
+i0^{+}\right)  \right]  _{24}\right\}  . \label{hipm}%
\end{align}
These functions are useful to distinguish between the contributions of the
total and relative responses to the analytic solutions of the dispersion
equation $\det\mathbb{M}_{2b}\left(  \mathbf{q},z\right)  =0$. An even more
clear identification of different modes is possible using spectral weight
functions for total and relative amplitude and phase responses. These spectral
weight functions are determined using the inverse GPF propagator in the basis
of the amplitude and phase field coordinates. The inverse GPF propagator for a
one-band system is determined as in Ref. \cite{Engelbrecht}:%
\begin{equation}
\mathbb{Q}\left(  \mathbf{q},z\right)  =\left(
\begin{array}
[c]{cc}%
Q_{1,1}\left(  \mathbf{q},z\right)  & Q_{1,2}\left(  \mathbf{q},z\right) \\
Q_{2,1}\left(  \mathbf{q},z\right)  & Q_{2,2}\left(  \mathbf{q},z\right)
\end{array}
\right)
\end{equation}
with the matrix elements:%
\begin{align}
Q_{1,1}\left(  \mathbf{q},z\right)   &  =M_{1,1}^{\left(  E\right)  }\left(
\mathbf{q},z\right)  +M_{12}\left(  \mathbf{q},z\right)  ,\\
Q_{2,2}\left(  \mathbf{q},z\right)   &  =M_{1,1}^{\left(  E\right)  }\left(
\mathbf{q},z\right)  -M_{12}\left(  \mathbf{q},z\right)  ,\\
Q_{1,2}\left(  \mathbf{q},z\right)   &  =iM_{1,1}^{\left(  A\right)  }\left(
\mathbf{q},z\right)  ,\\
Q_{2,1}\left(  \mathbf{q},z\right)   &  =-iM_{1,1}^{\left(  A\right)  }\left(
\mathbf{q},z\right)  ,
\end{align}
and%
\begin{align}
M_{1,1}^{\left(  E\right)  }\left(  \mathbf{q},z\right)   &  =\frac
{M_{1,1}\left(  \mathbf{q},z\right)  +M_{1,1}\left(  \mathbf{q},-z\right)
}{2},\\
M_{1,1}^{\left(  A\right)  }\left(  \mathbf{q},z\right)   &  =\frac
{M_{1,1}\left(  \mathbf{q},z\right)  -M_{1,1}\left(  \mathbf{q},-z\right)
}{2}.
\end{align}
In this basis, the diagonal matrix elements $Q_{1,1}$ and $Q_{2,2}$ correspond
to amplitude and phase collective excitations, respectively. The non-diagonal
matrix elements describe their mixing. Obviously, $\det\mathbb{Q=}%
\det\mathbb{M}$.

For a two-band system, the inverse GPF propagator matrix in the basis of the
amplitude and phase field coordinates is obtained straightforwardly:%
\begin{equation}
\mathbb{Q}_{2b}=\left(
\begin{array}
[c]{cc}%
\mathbb{\tilde{Q}}^{\left(  1\right)  } & -\varkappa\mathbb{I}\\
-\varkappa\mathbb{I} & \mathbb{\tilde{Q}}^{\left(  2\right)  }%
\end{array}
\right)  ,
\end{equation}
with the notations
\begin{align}
\mathbb{\tilde{Q}}^{\left(  1\right)  }  &  =\mathbb{Q}^{\left(  1\right)
}+\varkappa\frac{\Delta_{2}}{\Delta_{1}}\cdot\mathbb{I},\\
\mathbb{\tilde{Q}}^{\left(  2\right)  }  &  =\mathbb{\tilde{Q}}^{\left(
2\right)  }+\varkappa\frac{\Delta_{1}}{\Delta_{2}}\cdot\mathbb{I}.
\end{align}
The GPF propagator $\mathbb{\tilde{Q}}_{2b}^{-1}$ in this basis is explicitly
given by:
%TCIMACRO{\TeXButton{Begin widetext}{\begin{widetext}}}%
%BeginExpansion
\begin{widetext}%
%EndExpansion%
\begin{equation}
\mathbb{\tilde{Q}}_{2b}^{-1}=\left(
\begin{array}
[c]{cccc}%
\frac{\tilde{Q}_{22}^{\left(  1\right)  }\det\mathbb{\tilde{Q}}^{\left(
2\right)  }-\varkappa^{2}\tilde{Q}_{11}^{\left(  2\right)  }}{\det
\mathbb{\tilde{Q}}_{2b}} & -\frac{\tilde{Q}_{12}^{\left(  1\right)  }%
\det\mathbb{\tilde{Q}}^{\left(  2\right)  }+\varkappa^{2}\tilde{Q}%
_{12}^{\left(  2\right)  }}{\det\mathbb{\tilde{Q}}_{2b}} & \varkappa
\frac{\tilde{Q}_{22}^{\left(  1\right)  }\tilde{Q}_{22}^{\left(  2\right)
}-\tilde{Q}_{12}^{\left(  1\right)  }\tilde{Q}_{12}^{\left(  2\right)
}-\varkappa^{2}}{\det\mathbb{\tilde{Q}}_{2b}} & -\varkappa\frac{\tilde{Q}%
_{12}^{\left(  1\right)  }\tilde{Q}_{11}^{\left(  2\right)  }+\tilde{Q}%
_{22}^{\left(  1\right)  }\tilde{Q}_{12}^{\left(  2\right)  }}{\det
\mathbb{\tilde{Q}}_{2b}}\\
\frac{\tilde{Q}_{12}^{\left(  1\right)  }\det\mathbb{\tilde{Q}}^{\left(
2\right)  }+\varkappa^{2}\tilde{Q}_{12}^{\left(  2\right)  }}{\det
\mathbb{\tilde{Q}}_{2b}} & \frac{\tilde{Q}_{11}^{\left(  1\right)  }%
\det\mathbb{\tilde{Q}}^{\left(  2\right)  }-\varkappa^{2}\tilde{Q}%
_{22}^{\left(  2\right)  }}{\det\mathbb{\tilde{Q}}_{2b}} & \varkappa
\frac{\tilde{Q}_{11}^{\left(  1\right)  }\tilde{Q}_{12}^{\left(  2\right)
}+\tilde{Q}_{12}^{\left(  1\right)  }\tilde{Q}_{22}^{\left(  2\right)  }}%
{\det\mathbb{\tilde{Q}}_{2b}} & \varkappa\frac{\tilde{Q}_{11}^{\left(
1\right)  }\tilde{Q}_{11}^{\left(  2\right)  }-\tilde{Q}_{12}^{\left(
1\right)  }\tilde{Q}_{12}^{\left(  2\right)  }-\varkappa^{2}}{\det
\mathbb{\tilde{Q}}_{2b}}\\
\varkappa\frac{\tilde{Q}_{22}^{\left(  1\right)  }\tilde{Q}_{22}^{\left(
2\right)  }-\tilde{Q}_{12}^{\left(  1\right)  }\tilde{Q}_{12}^{\left(
2\right)  }-\varkappa^{2}}{\det\mathbb{\tilde{Q}}_{2b}} & -\varkappa
\frac{\tilde{Q}_{11}^{\left(  1\right)  }\tilde{Q}_{12}^{\left(  2\right)
}+\tilde{Q}_{12}^{\left(  1\right)  }\tilde{Q}_{22}^{\left(  2\right)  }}%
{\det\mathbb{\tilde{Q}}_{2b}} & \frac{\tilde{Q}_{22}^{\left(  2\right)  }%
\det\mathbb{\tilde{Q}}^{\left(  1\right)  }-\tilde{Q}_{11}^{\left(  1\right)
}\varkappa^{2}}{\det\mathbb{\tilde{Q}}_{2b}} & -\frac{\tilde{Q}_{12}^{\left(
2\right)  }\det\mathbb{\tilde{Q}}^{\left(  1\right)  }+\tilde{Q}_{12}^{\left(
1\right)  }\varkappa^{2}}{\det\mathbb{\tilde{Q}}_{2b}}\\
\varkappa\frac{\tilde{Q}_{12}^{\left(  1\right)  }\tilde{Q}_{11}^{\left(
2\right)  }+\tilde{Q}_{22}^{\left(  1\right)  }\tilde{Q}_{12}^{\left(
2\right)  }}{\det\mathbb{\tilde{Q}}_{2b}} & \varkappa\frac{\tilde{Q}%
_{11}^{\left(  1\right)  }\tilde{Q}_{11}^{\left(  2\right)  }+\tilde{Q}%
_{12}^{\left(  1\right)  }\tilde{Q}_{12}^{\left(  2\right)  }-\varkappa^{2}%
}{\det\mathbb{\tilde{Q}}_{2b}} & \frac{\tilde{Q}_{12}^{\left(  2\right)  }%
\det\mathbb{\tilde{Q}}^{\left(  1\right)  }+\tilde{Q}_{12}^{\left(  1\right)
}\varkappa^{2}}{\det\mathbb{\tilde{Q}}_{2b}} & \frac{\tilde{Q}_{11}^{\left(
2\right)  }\det\mathbb{\tilde{Q}}^{\left(  1\right)  }-\varkappa^{2}\tilde
{Q}_{22}^{\left(  1\right)  }}{\det\mathbb{\tilde{Q}}_{2b}}%
\end{array}
\right)  .
\end{equation}

In the same way as in the above subsection, we introduce four spectral weight
functions, which describe total $\left(  +\right)  $ and relative $\left(
-\right)  $ amplitude-amplitude ($aa$) and phase ($pp$) pair field responses:%
\begin{align}
\chi_{aa}^{\left(  \pm\right)  }\left(  \mathbf{q},\omega\right)   &
=\frac{1}{\pi}\operatorname{Im}\left\{  \left[  \mathbb{\tilde{Q}}_{2b}%
^{-1}\left(  \mathbf{q},\omega+i0^{+}\right)  \right]  _{11}+\left[
\mathbb{\tilde{Q}}_{2b}^{-1}\left(  \mathbf{q},\omega+i0^{+}\right)  \right]
_{33}\pm2\left[  \mathbb{\tilde{Q}}_{2b}^{-1}\left(  \mathbf{q},\omega
+i0^{+}\right)  \right]  _{13}\right\}  ,\label{hiaa}\\
\chi_{pp}^{\left(  \pm\right)  }\left(  \mathbf{q},\omega\right)   &
=\frac{1}{\pi}\operatorname{Im}\left\{  \left[  \mathbb{\tilde{Q}}_{2b}%
^{-1}\left(  \mathbf{q},\omega+i0^{+}\right)  \right]  _{22}+\left[
\mathbb{\tilde{Q}}_{2b}^{-1}\left(  \mathbf{q},\omega+i0^{+}\right)  \right]
_{44}\pm2\left[  \mathbb{\tilde{Q}}_{2b}^{-1}\left(  \mathbf{q},\omega
+i0^{+}\right)  \right]  _{24}\right\}  . \label{hipp}%
\end{align}%
%TCIMACRO{\TeXButton{End widetext}{\end{widetext}}}%
%BeginExpansion
\end{widetext}%
%EndExpansion


\begin{thebibliography}{99}                                                                                               %


\bibitem {Leggett}A. J. Leggett, \textquotedblleft\emph{Number-Phase
Fluctuations in Two-Band Superconductors}\textquotedblright, Progr. Theor.
Phys. \textbf{36}, 901 (1966).

\bibitem {Blumberg}G. Blumberg, A. Mialitsin, B. \thinspace S. Dennis, M. V.
Klein, N. D. Zhigadlo, and J. Karpinski, \textquotedblleft\emph{Observation of
Leggett's Collective Mode in a Multiband MgB}$_{2}$ \emph{Superconductor}%
\textquotedblright, Phys. Rev. Lett. \textbf{99}, 227002 (2007).

\bibitem {Ponomarev}Ya. G. Ponomarev, S. A. Kuzmichev, M. G. Mikheev, M. V.
Sudakova, S. N. Tchesnokov, N. Z. Timergaleev, A. V. Yarigin, E. G. Maksimov,
S. I. Krasnosvobodtsev, A.V.Varlashkin, M. A. Hein, G.M\"{u}ller, H. Piel, L.
G. Sevastyanova, O. V. Kravchenko, K. P. Burdina, and B. M. Bulychev,
\textquotedblleft\emph{Evidence for a two-band behavior of MgB}$_{2}$
\emph{from point-contact and tunneling spectroscopy}\textquotedblright, Solid
State Commun. \textbf{129}, 85 (2004).

\bibitem {Giorgianni}F. Giorgianni, T. Cea, C. Vicario, C. P. Hauri, W. K.
Withanage, X Xi and L. Benfatto, \textquotedblleft\emph{Leggett mode
controlled by light pulses}\textquotedblright, Nature Phys. \textbf{15}, 341 (2019).

\bibitem {Maksimov}E. G. Maksimov, A. E. Karakozov, B. P. Gorshunov, Ya. G.
Ponomarev, E. S. Zhukova, and M. Dressel, \textquotedblleft\emph{Theoretical
analysis of two-gap superconductivity of magnesium diborides and iron
pnictides in the generalized }$\alpha$ \emph{model}\textquotedblright, JETP
\textbf{115}, 252 (2012).

\bibitem {Iskin1}M. Iskin and C. A. R. S\'{a} de Melo, \textquotedblleft%
\emph{BCS-BEC crossover of collective excitations in two-band superfluids}%
\textquotedblright, Phys. Rev. B \textbf{72}, 024512 (2005).

\bibitem {Iskin2}M. Iskin and C. A. R. S\'{a} de Melo, \textquotedblleft%
\emph{Two-band superfluidity from the BCS to the BEC limit}\textquotedblright,
Phys. Rev. B \textbf{74}, 144517 (2006).

\bibitem {Zhang2015}R. Zhang, Y. Cheng, H. Zhai, and P. Zhang,
\textquotedblleft\emph{Orbital Feshbach Resonance in Alkali-Earth
Atoms}\textquotedblright, Phys. Rev. Lett. \textbf{115}, 135301 (2015).

\bibitem {Pagano}G. Pagano, M. Mancini, G. Cappellini, L. Livi, C. Sias, J.
Catani, M. Inguscio, and L. Fallani, \textquotedblleft\emph{Strongly
Interacting Gas of Two-Electron Fermions at an Orbital Feshbach Resonance}%
\textquotedblright, Phys. Rev. Lett. \textbf{115}, 265301 (2015).

\bibitem {Hofer}M. H\"{o}fer, L. Riegger, F. Scazza, C. Hofrichter,
D.\thinspace R. Fernandes, M.\thinspace M. Parish, J. Levinsen, I. Bloch, and
S. F\"{o}lling, \textquotedblleft\emph{Observation of an Orbital
Interaction-Induced Feshbach Resonance in }$^{173}$\emph{Yb}\textquotedblright%
, Phys. Rev. Lett. \textbf{115}, 265302 (2015).

\bibitem {Iskin3}M. Iskin, \textquotedblleft\emph{Two-band superfluidity and
intrinsic Josephson effect in alkaline-earth-metal Fermi gases across an
orbital Feshbach resonance}\textquotedblright, Phys. Rev. A \textbf{94},
011604 (2016).

\bibitem {He2015}L. He, H. Hu, and X.-J. Liu, \textquotedblleft\emph{Two-band
description of resonant superfluidity in atomic Fermi gases}\textquotedblright%
, Phys. Rev. A \textbf{91}, 023622 (2015).

\bibitem {He2016}L. He, J. Wang, S.-G. Peng, X.-J. Liu, and H. Hu,
\textquotedblleft\emph{Strongly correlated Fermi superfluid near an orbital
Feshbach resonance: Stability, equation of state, and Leggett mode}%
\textquotedblright, Phys. Rev. A \textbf{94}, 043624 (2016).

\bibitem {Zhang}Y.-C. Zhang, S. Ding, and S. Zhang, \textquotedblleft%
\emph{Collective modes in a two-band superfluid of ultracold
alkaline-earth-metal atoms close to an orbital Feshbach resonance}%
\textquotedblright, Phys. Rev. A \textbf{95}, 041603(R) (2017).

\bibitem {Tajima2019}H. Tajima, Y. Yerin, A. Perali, and P. Pieri,
\textquotedblleft\emph{Enhanced critical temperature, pairing fluctuation
effects, and BCS-BEC crossover in a two-band Fermi gas}\textquotedblright,
Phys. Rev. B \textbf{99}, 180503(R) (2019).

\bibitem {Yerin2019}Y. Yerin, H. Tajima, P. Pieri, and A. Perali,
\textquotedblleft\emph{Coexistence of giant Cooper pairs with a bosonic
condensate and anomalous behavior of energy gaps in the BCS-BEC crossover of a
two-band superfluid Fermi gas}\textquotedblright, Phys. Rev. B \textbf{100},
104528 (2019).

\bibitem {Zou2018}P. Zou, L. He, X.-J. Liu, and H. Hu, \textquotedblleft%
\emph{Strongly interacting Sarma superfluid near orbital Feshbach
resonances}\textquotedblright, Phys. Rev. A \textbf{97}, 043616 (2018).

\bibitem {Klimin2015}S. N. Klimin, J. Tempere, G. Lombardi, and J. T.
Devreese, \textquotedblleft\emph{Finite temperature effective field theory and
two-band superfluidity in Fermi gases}\textquotedblright, Eur. Phys. Journal B
\textbf{88}, 122 (2015).

\bibitem {Kurkjian2019}H. Kurkjian, S. N. Klimin, J. Tempere, and Y. Castin,
\textquotedblleft\emph{Pair-breaking collective branch in BCS superconductors
and superfluid Fermi gases}\textquotedblright, Phys. Rev. Lett. \textbf{122},
093403 (2019).

\bibitem {Arxiv}S. N. Klimin, J. Tempere, and H. Kurkjian,\textquotedblleft%
\emph{Phononic collective excitations in superfluid Fermi gases at nonzero
temperatures}\textquotedblright, arXiv:1811.07796 (2019).

\bibitem {Nozieres}P. Nozi\`{e}res, \textquotedblleft\emph{Le probleme a }$N$
\emph{corps: propri\'{e}t\'{e}s g\'{e}n\'{e}rales des gaz de fermions}%
\textquotedblright\ (Dunod, Paris, 1963).

\bibitem {SDM1993}C. A. R. S\'{a} de Melo, M. Randeria, and J.R. Engelbrecht,
\textquotedblleft\emph{Crossover from BCS to Bose superconductivity:
Transition temperature and time-dependent Ginzburg-Landau theory}%
\textquotedblright, Phys. Rev. Lett. \textbf{71}, 3202 (1993).

\bibitem {Engelbrecht}J. R. Engelbrecht, M. Randeria, and C. A. R. S\'{a} de
Melo, \textquotedblleft\emph{BCS to Bose crossover: Broken-symmetry
state}\textquotedblright, Phys. Rev. B \textbf{55}, 15153 (1997).

\bibitem {Diener2008}R. B. Diener, R. Sensarma, and M. Randeria,
\textquotedblleft\emph{Quantum fluctuations in the superfluid state of the
BCS-BEC crossover}\textquotedblright, Phys. Rev. A \textbf{77}, 023626 (2008).

\bibitem {PRA2008}J. Tempere, S. N. Klimin, and J. T. Devreese,
\textquotedblleft\emph{Phase separation in imbalanced fermion superfluids
beyond the mean-field approximation}\textquotedblright, Phys. Rev. A
\textbf{78}, 023626 (2008).

\bibitem {Anderson}P. W. Anderson, \textquotedblleft\textit{Random-Phase
Approximation in the Theory of Superconductivity}\textquotedblright, Phys.
Rev. \textbf{112}, 1900 (1958).

\bibitem {Han}X. Han, B. Liu, and J. Hu, \textquotedblleft\emph{Observability
of Higgs mode in a system without Lorentz invariance}\textquotedblright, Phys.
Rev. A \textbf{94}, 033608 (2016).

\bibitem {Deng}T.-S. Deng, W. Zhang, and W. Yi, \textquotedblleft\emph{Tuning
Feshbach resonances in cold atomic gases with interchannel coupling}%
\textquotedblright, Phys. Rev. A \textbf{96}, 050701(R) (2017).

\bibitem {Castin2001}A.~Minguzzi, G.~Ferrari, and Y.~Castin, \textquotedblleft%
\emph{Dynamic structure factor of a superfluid Fermi gas}\textquotedblright,
Eur. Phys. Journal D, \textbf{17}, 49 (2001).
\end{thebibliography}
\end{document}